\documentclass[11pt,preprintnumbers,aps,amssymb,nofootinbib,amsmath,superscriptaddress]
{revtex4}
\usepackage{epsfig,epsf}
\usepackage{bm} 
\newcommand{\beq}{\begin{equation}}
\newcommand{\beql}[1]{\begin{equation}\label{#1}}
\newcommand{\eeq}{\end{equation}}
\def\bal#1\gal{\begin{align}#1\end{align}}
\newcommand{\ball}[1]{\bal\label{#1}}
\newcommand{\eq}[1]{(\ref{#1})}
\newcommand{\fig}[1]{Fig.~\ref{#1}}
\renewcommand{\sec}[1]{Sec.~\ref{#1}}
\newcounter{topiccounter}
\renewcommand{\b}[1]{{\bm #1}} 
\newcommand{\unit}[1]{\hat {{\bm #1}}} 

\newcommand{\e}{\varepsilon}
\newcommand{\aver}[1]{\left\langle #1 \right\rangle}

\begin{document}


\title{Synchrotron radiation of vector bosons  at relativistic colliders }

\author{Kirill Tuchin}

\affiliation{Department of Physics and Astronomy, Iowa State University, Ames, IA 50011}

\date{\today}


\begin{abstract}

Magnetic fields produced in collisions of electrically charged particles at relativistic energies are strong enough to affect the dynamics of the strong interactions. In particular, it induces radiation of vector bosons by relativistic fermions. I calculate the corresponding spectrum in constant magnetic field and analyze its angular distribution and mass dependence. As an application, synchrotron radiation of vector bosons by relativistic plasmas is considered.

\end{abstract}

\maketitle

\section{Introduction}\label{sec:intr}

It has been known since the pioneering paper of Ambjorn and Olesen \cite{Ambjorn:1990jg} that extremely strong electromagnetic fields are produced in high energy  collisions of charged particles. In recent years it was realized that these fields have an important impact on the dynamics of the strong interactions, though their precise structure and dynamics is being debated \cite{Kharzeev:2007jp,Tuchin:2013ie,McLerran:2013hla,Bzdak:2011yy,Gursoy:2014aka,Zakharov:2014dia}. In this paper we focus on vector boson radiation by relativistic particles in an external magnetic field. In particular, we are interested in real and virtual photon production, which  has important applications to the phenomenology of heavy-ion collisions \cite{Tuchin:2012mf,Tuchin:2010gx,Tuchin:2013bda}, astrophysics \cite{Harding:2006qn} and the physics of intense laser pulses  \cite{laser}.

The real photon radiation rate in vacuum was calculated in \cite{Sokolov:1968a} and is given by an infinite sum over the Landau levels. Based on this result synchrotron radiation from electromagnetic plasmas was calculated in \cite{Baring:1988a,Herold:1982a,Harding:1987a}. Pair production by a photon in an external magnetic field is a cross channel of the synchrotron radiation. The most general expression for the pair production probability by a virtual photon in vacuum is derived in \cite{Baier:2007dw}. The results of \cite{Sokolov:1968a,Baier:2007dw} are especially useful in very strong fields (defined below) when only a few lowest Landau levels contribute to the radiation rate. At not so strong fields and at ultra-relativistic energies, summation over the Landau levels is slowly convergent and is not convenient to deal with (in the context of heavy-ion physics see \cite{Tuchin:2012mf} for a detailed discussion of this issue). An alternative efficient method to calculate the scattering matrix in the ultra-relativistic approximation was developed by Baier and Katkov (see e.g.\ \cite{Baier-book} and references therein) and is described in  \cite{Berestetsky:1982aq}. It is based on the quasi-classical approximation and allows one to perform explicit summation over the Landau levels yielding rather simple formulas that are convenient in numerical and analytical calculations, see e.g.\ \cite{Baring:1988a}. While synchrotron radiation of real photons is of great interest in astrophysics, radiation of massive vectors bosons is of interest in heavy-ion collisions and in high intensity laser physics. Thus, in view of possible applications, it is very useful to have a compact expression for the synchrotron radiation of vector bosons. The goal of this paper is to feel the gap in the literature by calculating the synchrotron radiation of massive vector bosons and in particular virtual photons using the quasi-classical method.   

In order to calculate the vector meson production rate we need to know their coupling to quarks. A simple model inspired by the Vector Meson Dominance is to assume that coupling of different vector mesons to quarks has the same structure as the coupling of the photon. The corresponding terms in the Lagrangian are
\ball{z11}
\mathcal{L}_\gamma=e  \bar q \gamma^\mu Q q\, A_\mu\,,\quad
\mathcal{L}_\rho=g_\rho  \bar q \gamma^\mu \b \tau  q\cdot \b\rho_\mu\,,\quad
\mathcal{L}_\omega=g_\omega  \bar q \gamma^\mu  q\, \omega_\mu\,,
\gal
where $q$ is the SU(2) doublet of $u$ and $d$ quarks, $\b \tau$ are symmetry generators and $Q=\text{diag}(q_u,q_d)$. Eqs.~\eq{z11} constitute a part of the quark--meson coupling model \cite{Guichon:1987jp,Guichon:1995ue}, which is used to describe the nuclear matter. Similar approach is successfully used for calculation of the vector meson production at high energy in perturbative QCD  \cite{Nemchik:1996cw,Kopeliovich:2001xj}.

Throughout the paper we employ the ultra-relativistic approximation that requires fermion  and the vector boson to be relativistic and assume that magnetic field is adiabatic. Let $p=(\e,\b p)$ be the initial fermion four-momentum and $k= (\omega,\b k)$ the vector boson four-momentum, $m$ and $M$ their respective masses. Ultra-relativistic approximation requires that fermion energy before and after the vector boson emission satisfy $\e\gg m$ and $\e'= \e-\omega \gg m$. This implies that $\e'/\e\gg m/\e$ meaning that the  vector boson does not carry away all the fermion energy. Another implication of the ultra-relativistic approximation, which is instrumental for the spectrum derivation in the next section, is that the angular distribution of the vector boson spectrum is concentrated inside a narrow solid angle with the opening angle $\theta$ around the fermion direction. This can be seen by examining the denominator of the  outgoing fermion propagator  
\ball{a21}
(p-k)^2-m^2\approx -\e\omega\left( \frac{m^2}{\e^2}+\frac{M^2}{\omega^2}\frac{\e'}{\e}+\theta^2\right)\,.
\gal
The same expression appears in the argument of the Airy function in the formulas for the spectrum \eq{b77},\eq{b78}. Thus, the radiation cone  is determined by the largest among the small ratios $m/\e$ and  $\sqrt{M^2\e'}/\sqrt{\omega^2\e}< M/\omega$.  

The distance between the energy levels of a fermion in magnetic field is of the order of  $eB/\e$. If $eB\ll \e^2$ the spectrum can be considered as approximately continuous. This is always true in fields weaker than the Schwinger field $B_S=m^2/e$. In the following I will assume that the magnetic field strength is such that the quasi-classical approximation holds, i.e.\ $eB\ll \e^2$  (but not necessarily  $B<B_S$).

 The paper is structured as follows: In \sec{sec:b-2} I derive the vector boson spectrum radiated by a fast fermion moving in a plane perpendicular to the direction of magnetic field and in \sec{sec:b-3} I  analyze its mass dependence. In \sec{sec:d} the spectrum is boosted to an arbitrary frame. Sec.~\ref{sec:d} is dedicated to synchrotron radiation from plasma. Conclusions are presented in \sec{sec:con}.

\section{Vector boson radiation in the reaction plane $\b p\cdot \b B=0$.}\label{sec:b}

\subsection{Calculation of the  spectrum }\label{sec:b-2}

For the calculation of the vector boson spectrum I employ the method  described in \cite{Baier-book,Berestetsky:1982aq}. I follow the notations  of \cite{Berestetsky:1982aq} apart from minor changes. The calculation  is convenient to do in the frame $K_0$  where the fermion's momentum is perpendicular to the direction of magnetic field. The emission probability per unit time reads \cite{Berestetsky:1982aq}
\ball{b11}
d \dot w= \frac{\alpha}{(2\pi)^2}\frac{d^3k}{\omega}\int_{-\infty}^\infty d\tau \aver{R^*_2R_1}e^{i\Phi}\,,
\gal
where $\alpha= g^2/4\pi$ ($g$ stands for $e$, $g_\rho$, or $g_\omega$), $\aver{R^*_2R_1}$ denotes the average over the initial fermion polarization and summation over the final fermion and boson polarization and 
\bal
\Phi&= \frac{\e}{\e'}[\b k\cdot \b r_2-\b k\cdot \b r_1-\omega \tau]+\frac{M^2\tau}{2\e'} \,,\label{b12}\\
R&=- \frac{\bar u(p')}{\sqrt{2\e'}}\,\gamma\cdot \epsilon^*\,\frac{u(p)}{\sqrt{2\e}} \,. \label{b15}
\gal
The indexes 1 and 2 is a shorthand notation meaning that the corresponding quantity is taken at time $t_1=t+\tau/2$ or $t_2=t-\tau/2$. The bi-spinor is normalized as follows:
\ball{b13}
u(p)= \frac{1}{\sqrt{\e+m}}\left(\begin{array}{c}(\e+m)\varphi_p \\(\b p\cdot \b \sigma)\varphi_p\end{array}\right)\,,
\gal
where $\varphi_p$ is a two-component spinor and $\b\sigma$ are Pauli matrices.  The four-momentum of the incident fermion can be written as $p= \e(1,\b v)$. Similarly, I denote 
\ball{b17}
s=\sqrt{1-\frac{M^2}{\omega^2}}\,, 
\gal
so that the vector boson four-momentum can be written as $k=(\omega,\b k)= \omega(1, s \b n)$, where $\b n$ is a unit vector. Substituting \eq{b13} into \eq{b15} I obtain for transversely polarized boson 
\ball{b19}
 R_T= \varphi_{p'}^*\b\epsilon_T^*\cdot (\b A+i\b B\times \b \sigma)\varphi_{p}\,,
\gal
where the following auxiliary vectors are introduced:
\bal
\b A &= \left( \sqrt{\frac{\e'+m}{\e+m}}+\sqrt{\frac{\e+m}{\e'+m}}\right)\frac{\sqrt{\e}}{\sqrt{2\e'}}\b v\,,\label{b21}\\
\b B & = \left[\left( \sqrt{\frac{\e'+m}{\e+m}}-\sqrt{\frac{\e+m}{\e'+m}}\right)\b p +\sqrt{\frac{\e+m}{\e'+m}}\,\b k\right]\frac{1}{2\sqrt{\e\e'}}\,,\label{b22}
\gal
and $\e'=\e-\omega$.
Multiplying \eq{b19} by its complex conjugate and averaging using the formula $\aver{\epsilon_{T,j} \epsilon_{T,k}}= (\delta_{jk}-n_jn_k)/2$ we get
\ball{b23}
\aver{R^*_{T,2}R_{T,1}}= \b A_1\cdot \b A_2-(\b A_1\cdot \b n)(\b A_2\cdot \b n)+\b B_1\cdot \b B_2+(\b B_1\cdot \b n)(\b B_2\cdot \b n)\,.
\gal 
Expanding \eq{b21},\eq{b22} in $m^2/\e^2$ and $M^2/\omega^2$ yields
\bal
&\b A\approx \left(1+\frac{\e}{\e'}\right)\frac{\b v}{2}\,,\label{b25}\\
&\b B\approx \frac{\omega}{2\e'}\left(-\b v+\b n+\frac{m}{\e}\b n+(s-1)\b n\right)\,.\label{b26}
\gal
The terms like $\b v_1\cdot \b n$ arising in \eq{b23}  can be simplified  using  integration by parts in \eq{b11} as follows \cite{Berestetsky:1982aq}
\ball{b28}
\b v_1\cdot \b n\, e^{i\Phi} =\b v_2\cdot \b n\, e^{i\Phi} = \frac{1}{s}\left[ 1+\frac{\omega (s^2-1)}{2\e }\right] e^{i\Phi}  \,,
\gal
where  the terms proportional to the total time derivative with respect to  $t_1$, which vanish upon integration over time in \eq{b11}, are dropped.  Substituting \eq{b25},\eq{b26} into \eq{b23}   I derive
\ball{b31}
 \aver{R^*_{T,2}R_{T,1}}=\frac{\e'^2+\e^2}{2\e'^2}(\b v_1\cdot \b v_2-1)-\frac{M^2}{2\omega^2}\left( \frac{\e}{\e'}+\frac{\e'}{\e}\right)+\frac{\omega^2m^2}{2\e^2\e'^2}\,.
\gal 
The explicit expression for the fermion trajectory in a plane perpendicular to magnetic field yields at small $\tau$:
\ball{b33}
\b v_1\cdot \b v_2= 1-\frac{m^2}{\e^2}-\frac{1}{2}\omega_B^2\tau^2\,, 
\gal
where $\omega_B= eB/\e$ is the synchrotron frequency. Thus, \eq{b31} takes form
\ball{b35}
 \aver{R^*_{T,2}R_{T,1}}=-\frac{\e'^2+\e^2}{4\e'^2}\omega_B^2\tau^2-\frac{M^2}{2\omega^2}\left( \frac{\e}{\e'}+\frac{\e'}{\e}\right)-\frac{m^2}{\e\e'}\,.
\gal

The longitudinal polarization is described by the four-vector $\epsilon_L= (s,\unit n)/\sqrt{s^2-1}$, which satisfies $\epsilon\cdot k=0$ and $\epsilon^2=1$. Writing $R= -j\cdot \epsilon$ and using the Ward identity $j\cdot k=0$ we have  $j^0=s\b j\cdot \b n$ implying that 
 \ball{b41}
 j\cdot \epsilon_L= \frac{j^0s-\b j\cdot \b n}{\sqrt{s^2-1}}= \sqrt{s^2-1}\,\b j\cdot \b n\,.
 \gal

Using \eq{b13} and \eq{b15} produces 
\ball{b45}
R_L= i\sqrt{1-s^2}\,\varphi_{p'}^*(F+ i\b \sigma\cdot \b G)\varphi_p\,.
\gal 
and
\ball{b47}
\aver{R^*_{L,2}R_{L,1}} = (1-s^2)(F_2F_1+\b G_2\cdot \b G_1)\,,
\gal
where 
\bal
F&=\frac{1}{2\sqrt{\e\e'}\sqrt{\e+m}\sqrt{\e'+m}}[(\e'+m)(\b n\cdot \b p)+(\e+m)(\b p'\cdot \b n)]\label{b49}\\
\b G& = \frac{1}{2\sqrt{\e\e'}\sqrt{\e+m}\sqrt{\e'+m}}[(\e'+m)(\b n\times \b p)+(\e+m)(\b n\times \b p')]\,,\label{b50}
\gal
with $\b p'= \b p-\b k$. In view of a small factor $1-s^2$ in the right hand side of \eq{b47} we only need  to keep terms of the order one in expansion of $F$ and $\b G$ in powers of $m^2/\e^2$ and $M^2/\omega^2$.
Thus, in view of \eq{b28}
$\b p\cdot \b n \approx \e$, $\b p'\cdot \b n \approx \e'$ and we have $F\approx 1$, $\b G\approx -\frac{\omega}{2\e'}\,\b n\times \b v$. This implies that  $\b G_1\cdot \b G_2\propto 1-\b v_1\cdot \b v_2\sim m^2/\e^2$ can be neglected and we derive 
\ball{b59}
\aver{R^*_{L,2}R_{L,1}} \approx \frac{M^2}{\omega^2}\,.
\gal

The expression in the exponent of \eq{b11} upon expansion  in $\tau$ and then in $M/\omega$ becomes
\ball{b71}
\Phi = -\frac{\e}{\e'}\omega \tau \left [1-s\b n\cdot \b v+\frac{\omega}{2\e}(s^2-1)+s\omega_B^2\frac{\tau^2}{24}\right]\approx -\frac{\e}{\e'}\omega \tau \left [1-\b n\cdot \b v+\frac{M^2}{2\omega^2}\frac{\e'}{\e}+\omega_B^2\frac{\tau^2}{24}\right]\,.
\gal
Substituting \eq{b35}, \eq{b59} and \eq{b71} into \eq{b11}  we obtain for the transverse and longitudinal  vector boson production rates 
\bal
d \dot w_T=& \frac{\alpha}{(2\pi)^2}\frac{d^3k}{\omega}\int_{-\infty}^\infty d\tau \exp\left\{-\frac{i\e}{\e'}\omega \tau \left [1-\b n\cdot \b v+\frac{M^2}{2\omega^2}\frac{\e'}{\e}+\omega_B^2\frac{\tau^2}{24}\right]\right\}\nonumber\\
&\times \,\left[ -\frac{\e'^2+\e^2}{4\e'^2}\omega_B^2\tau^2-\frac{M^2}{2\omega^2}\left( \frac{\e}{\e'}+\frac{\e'}{\e}\right)-\frac{m^2}{\e\e'}\right]\,,\label{b73} \\ 
d \dot w_L=& \frac{\alpha}{(2\pi)^2}\frac{d^3k}{\omega}\frac{M^2}{\omega^2}\int_{-\infty}^\infty d\tau \exp\left\{-\frac{i\e}{\e'}\omega \tau \left [1-\b n\cdot \b v+\frac{M^2}{2\omega^2}\frac{\e'}{\e}+\omega_B^2\frac{\tau^2}{24}\right]\right\}\,.\label{b74}
\gal
One can do the integrals over $\tau$  using equations \eq{ap11} and \eq{ap13} which yields the angular distribution of the spectrum
\bal
\frac{d \dot w_T}{d\omega d\Omega}= &\frac{\alpha}{\pi}\omega  \left(\frac{\e'}{\e \omega \omega_B^2} \right)^{1/3}
\left[ \frac{M^2}{2\omega^2}\frac{\e'^2+\e^2}{\e\e'}-\frac{m^2}{\e\e'}+2\left( 1-\b n\cdot \b v\right)\frac{\e'^2+\e^2}{\e'^2}\right] \nonumber\\
&\times \text{Ai}\left(  2\left( \frac{\e\omega}{\e'\omega_B}\right)^{2/3} \left( 1-\b n\cdot \b v+\frac{M^2\e'}{2\omega^2\e}\right) \right)\,,
\label{b77}\\
\frac{d \dot w_L}{d\omega d\Omega}= &\frac{\alpha}{\pi}\omega  \left(\frac{\e'}{\e \omega \omega_B^2} \right)^{1/3} \frac{M^2}{\omega^2}\,
\text{Ai}\left(  2\left( \frac{\e\omega}{\e'\omega_B}\right)^{2/3} \left( 1-\b n\cdot \b v+\frac{M^2\e'}{2\omega^2\e}\right) \right)\,,
\label{b78}
\gal
where we used $d^3k=s\omega^2d\omega d\Omega\approx \omega^2d\omega d\Omega$. Notice  the follwing expression  
\ball{b79}
2\left(1-\b n\cdot \b v+\frac{M^2\e'}{2\omega^2\e} \right)\approx  \theta^2+ \frac{m^2}{\omega^2}+\frac{M^2\e'}{2\omega^2\e}\,,
\gal
which appears in the argument of the Airy function. It is proportional to the denominator of the outgoing fermion propagator \eq{a21} and guarantees emission of vector boson into a narrow cone. 

The integration over the photon directions is convenient to do in \eq{b73},\eq{b74} followed by integration over $\tau$ \cite{Berestetsky:1982aq}. The result is 
\bal
\frac{d \dot w_T}{d\omega}= &-\frac{\alpha m^2}{\e^2}\left\{ \left( 1+\frac{M^2}{2\omega^2}\frac{\e^2+\e'^2}{m^2}\right)\int_z^\infty \text{Ai}(z')dz' + \left( \frac{\e}{\e'}\right)^{1/3} \left( \frac{\omega_B}{\omega}\right)^{2/3}\frac{\e^2+\e'^2}{m^2} \text{Ai}'(z)
\right\}\label{b81}\,,\\
\frac{d \dot w_L}{d\omega}= &\frac{\alpha M^2 \e'}{\omega^2\e}\int_z^\infty \text{Ai}(z')dz' \,,\label{b82}
\gal
where 
\ball{b84}
z=\left( \frac{\e}{\e'}\right)^{2/3} \left( \frac{\omega}{\omega_B}\right)^{2/3}\left( \frac{m^2}{\e^2}+\frac{M^2}{\omega^2}\frac{\e'}{\e}\right)\,.
\gal

\subsection{Analysis of the spectrum}\label{sec:b-3}

The vector boson spectrum \eq{b81},\eq{b82} is a function of  $\omega$ and  $\e$. Instead, we can express the spectrum in terms of the boost-invariant dimensionless quantities $X$ and $\xi$ defined as follows: 
\ball{b91}
X= \sqrt{-\frac{e^2}{m^6}\,(F_{\mu\nu}p^\nu)^2} \approx \frac{\omega_B\e^2}{m^3}= \frac{eB\e}{m^3}
\gal
and 
\ball{b93}
\xi = \frac{\omega}{\omega_c}\,,
\gal
where 
\ball{b95}
\omega_c= \frac{\e X}{\frac{2}{3}+X}\,,
\gal
is the characteristic frequency of the classical photon spectrum. Its is also convenient to denote $\mu= M/m$. In terms of these variables we can write  
\bal
&z= \frac{\xi^{2/3}}{[\frac{2}{3}+X(1-\xi)]^{2/3}}+\mu^2\frac{[\frac{2}{3}+X(1-\xi)]^{1/3}(\frac{2}{3}+X)}{X^2\xi^{4/3}}\,\label{b97}\,,\\
&\e'= \frac{\omega_c \left[\frac{2}{3}+X(1-\xi)\right]}{X}\,.\label{b98}
\gal
Becasue $\e'\ge 0$, it follows form \eq{b98} that
\ball{b100}
\xi\le \frac{2}{3X}+1\,.
\gal
 When multiplied by $\omega$, \eq{b81},\eq{b82} yield the radiation power. Dividing it by $3/2$ of the total classical photon radiation power $\alpha m^2X^2$  we represent the spectrum in terms of the dimensionless quantities
\ball{b102}
J_\lambda(\xi,X,\mu)= \frac{\omega}{\alpha m^2 X^2}\frac{d\dot w_\lambda}{d\xi}\,,\quad \lambda=L,T\,.
\gal
Their explicit form reads as follows
\bal
J_T=& -\frac{\xi}{(\frac{2}{3}+X)^2}\left\{ \left[ 1+\frac{\mu^2}{2X^2\xi^2}
\left( \frac{8}{9}+\frac{4}{3}X(2-\xi)+X^2(2-2\xi+\xi^2) \right)\right]
\int_z^\infty \text{Ai}(z')dz' \right.   \nonumber\\
&\quad
\left.  +
\frac{(\frac{2}{3}+X)^2+[\frac{2}{3}+X(1-\xi)]^2}{\xi^{2/3}[\frac{2}{3}+X(1-\xi)]^{1/3}(\frac{2}{3}+X)}
\text{Ai}'(z)\right\}
\,,\label{b104}\\
J_L=& \frac{\mu^2[\frac{2}{3}+X(1-\xi)]}{X^2\xi (\frac{2}{3}+X)}\int_z^\infty \text{Ai}(z')dz'\,. \label{b105}
\gal
 
The Airy function exponentially decays at large values of its argument, hence the spectrum is suppressed at $z\gg 1$. Variable $z$ as a function of $\xi$ has a minimum $z_0$ at $\xi_0$  that depends on the values of $X$ and $\mu$. The main contribution to the spectrum comes form the kinematic region $z<1$ which exists only if $z_0<1$. To determine $z_0$ and $\xi_0$ it is convenient to use instead of $\xi$ an auxiliary variable $u$:
\bal
u&= \frac{1}{\xi}\left[\frac{2}{3}+X(1-\xi)\right]\,,\label{b109}\\
 z&= \frac{1}{u^{2/3}}+\frac{\mu^2 u^{1/3}(u+X)}{X^2}\,.\label{b110}
\gal
The minimum of $z$ as a function of $u$ is located at 
\ball{b111}
u_0= \frac{X}{8}\left( \sqrt{1+\frac{32}{\mu^2}}-1\right)\,.
\gal
The corresponding value of $\xi$ reads
\ball{b114}
\xi_0= \frac{\frac{2}{3}+X}{u_0+X}= \frac{\frac{2}{3}+X}{X}\frac{8}{7+ \sqrt{1+\frac{32}{\mu^2}}}\,.
\gal
At $\mu\ll 1$, corresponding to an almost real photon,
\ball{b117}\
u_0\approx \frac{X}{\sqrt{2}\mu}\,,\quad \xi_0\approx \mu \sqrt{2}\,\frac{\frac{2}{3}+X}{X}\,,\quad \mu\ll 1\,.
\gal
Replacing $X= \sqrt{2}u_0\mu$ in \eq{b110} we get
\ball{b119}
z_0\approx \frac{3}{2u_0^{2/3}}\,,\quad \mu\ll 1\,.
\gal
Thus, the condition $z_0<1$ is satisfied only if $X>2.6\mu$. Otherwise, the spectrum is exponentially suppressed.  

In the opposite case, which is realized e.g.\ in production of high invariant mass dileptons,   $\mu^2\gg 32$ we have 
\ball{b119}
u_0\approx \frac{2X}{\mu^2}\,,\quad \xi_0\approx \frac{\frac{2}{3}+X}{X}\,,\quad \mu\gg 4\sqrt{2}\,, 
\gal
Comparing with \eq{b100}, we observe that in this case the minimum of $z$ is very close to the upper cutoff of the boson spectrum (i.e.\ when the boson takes nearly all energy of the fermion). 
Using $X=u_0\mu^2/2$ in \eq{b110} we have
\ball{b123}
z_0\approx \frac{3}{u_0^{2/3}}\,,\quad \mu\gg 4\sqrt{2}\,.
\gal
In this case $z_0<1$ is satisfied if $X> 2.6 \mu^2$ which is a much stronger condition than in the previous case. 

The main contribution to the spectrum arises from $z\sim 1$, which for $X$ and $\mu$ satisfying the above constraints and taking \eq{b110} into account  happens when $u\sim 1$ fairly independently from the value of $\mu$. This statement has been verified numerically.  In particular, according to \eq{b109} $u\sim 1$ means that $\xi\sim \frac{2}{3}+X(1-\xi)$.  In  weak fields $X\ll 1$, $\xi\sim 1$ and so $\omega\sim \omega_c\sim \e X$, while in strong fields $X\gg 1$, $\xi \sim X\e'/\omega_c$ (see \eq{b98}) implying that $\e'\sim \omega/X\sim \e/X\sim m^3/eB$ \cite{Berestetsky:1982aq}.

\begin{figure}[ht]
\begin{tabular}{cc}
      \includegraphics[height=4.5cm]{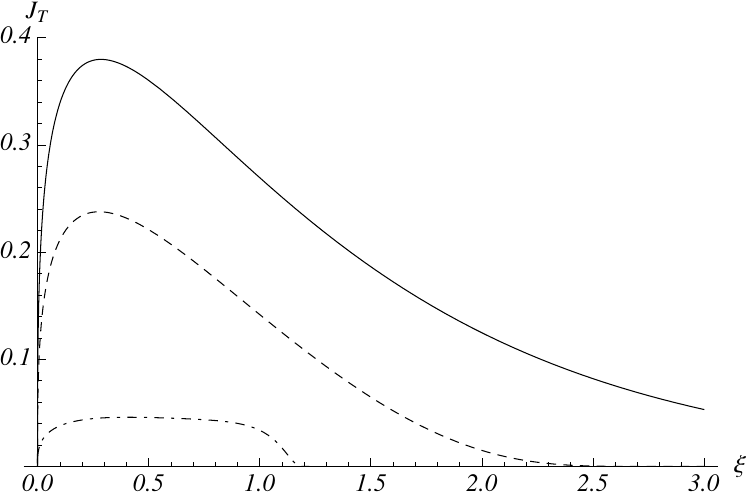} &
      \includegraphics[height=4.5cm]{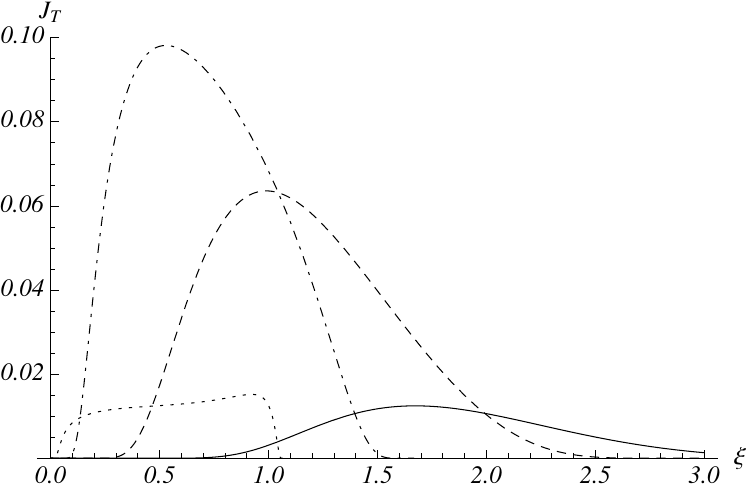}\\
      $(a)$ & $(b)$ \\
      \includegraphics[height=4.5cm]{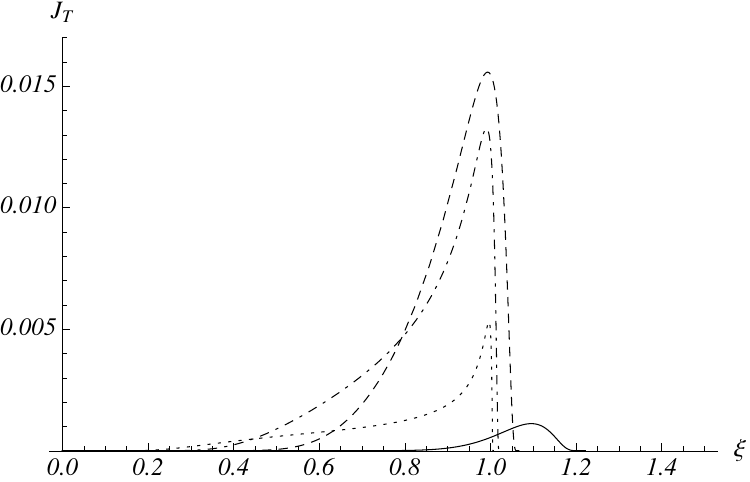} &
      \includegraphics[height=4.5cm]{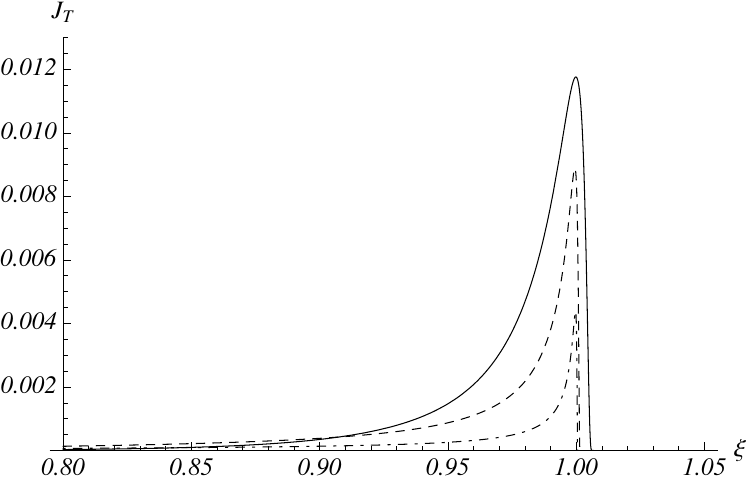}\\
      $(c)$ & $(d)$ 
      \end{tabular}
  \caption{Spectrum of transversely  polarized  vector bosons $J_T$ as a function of $\xi$. (a) $\mu=0$ and $X=0$ (solid line), $X=0.3$ (dashed line), $X=3$ (dash-dotted line). (b)  $\mu=0.3$ and $X=0.15$ (solid line), $X=0.3$ (dashed line), $X=1$ (dash-dotted line), $X=10$ (dotted lines). (c)  $\mu=3$ and $X=3$ (solid line), $X=10$ (dashed line), $X=30$ (dash-dotted line), $X=100$ (dotted lines). (d) $\mu=10$ and $X=100$ (solid line), $X=400$ (dashed line), $X=1000$ (dash-dotted line). Notice different scales of the $x$ and $y$ axes.}
\label{fig1}
\end{figure}

\begin{figure}[ht]
      \includegraphics[height=5cm]{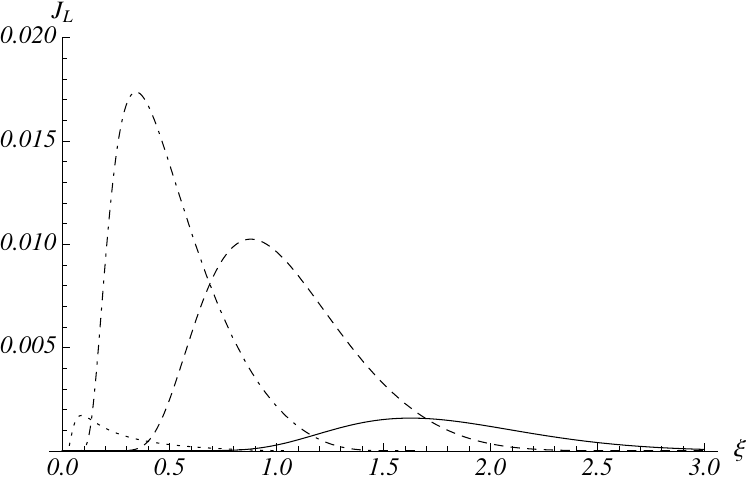} 
  \caption{Spectrum of longitudinaly  polarized  vector bosons $J_L$ as a function of $\xi$ at $\mu=0.3$ and $X=0.15$ (solid line), $X=0.3$ (dashed line), $X=1$ (dash-dotted line), $X=10$ (dotted lines)}
\label{fig2}
\end{figure}

\begin{figure}[ht]
      \includegraphics[height=5cm]{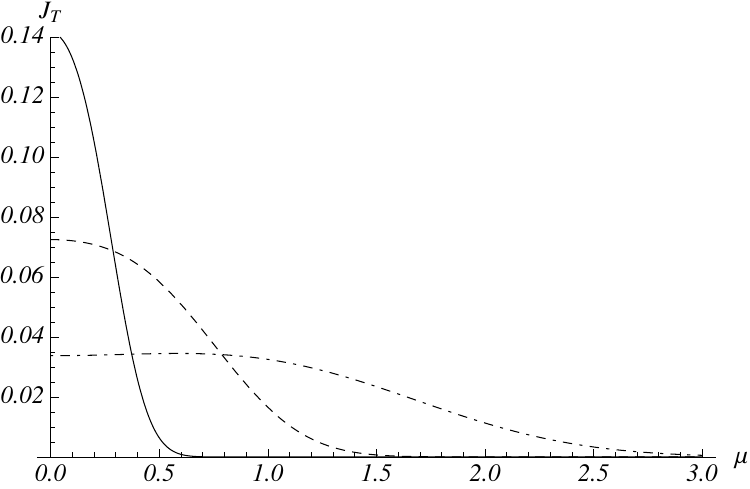} 
  \caption{Spectrum of transversely  polarized  vector bosons  $J_T$ as a function of $\mu$ at $\xi=1$ and $X=0.3$ (solid line), $X=1$ (dashed line) and $X=3$ (dashed-dotted line).}
\label{fig3}
\end{figure}

These features of the spectrum are seen in Figs.~\ref{fig1}--\ref{fig3}. In \fig{fig1}  the transverse vector boson spectrum as a function of $\xi$ is shown at different values of $X$ and $\mu$. The  transverse bosons are much more abundantly produced than the longitudinal ones, which can be seen by comparing \fig{fig1}(b) and \fig{fig2}. Therefore, \fig{fig1} represents approximately the total spectrum. The general trend observed in all figures is that the spectrum decreases with increase of $\mu$. At larger $\mu$ it tends to peak around $\xi=1$. This is because with increase of $\mu$, $X$ also increases, see the text after \eq{b119},\eq{b123}; it follows from \eq{b119} that  once $X\gg 1$, the typical $\xi$ is about 1.

\section{Vector boson spectrum in an arbitrary frame}\label{sec:d}

Consider now a reference frame $K$ where fermions have an arbitrary direction of momentum. It is convenient to change our notations. We will append a subscript 0 to all quantities pertaining to the reference frame $K_0$. Thus, for example, $\e_0$ and $\omega_0$  are the fermion and vector boson energies in $K_0$, whereas   $\e$ and $\omega$  are the fermion and vector boson energies in $K$.
Let the $y$-axis be in the magnetic field direction $\b B= B\unit y$ and $\b V = V\unit y$ be velocity of $K$ with respect to $K_0$. Then the Lorentz transformation reads
\bal
&p_{x0}= p_x\,, \quad 0=p_{y0}=  \gamma(p_y+V\e)\,,\quad p_{z0}= p_z\,,\quad \e_0= \gamma(\e+Vp_y)\,.\label{d11}\\
& k_{x0}= k_x\,,\quad k_{y0}= \gamma(k_y+V\omega)\,,\quad k_{z0}=k_z\,,\quad \omega_0= \gamma(\omega+Vk_y)\,.\label{d12}\\
&\b B_0= \b B\,, \label{d13}
\gal
where $\gamma= 1/\sqrt{1-V^2}$. It follows from the second equation in \eq{d11} that
\ball{d15}
V= -\frac{p_y}{\e}
\gal 
and 
\ball{d17}
\e_0= \sqrt{\e^2-p_y^2}\,,\qquad \omega_0= \frac{\omega \e- p_y k_y}{\sqrt{\e^2-p_y^2}}\,.
\gal
  Using the boost invariance of $k\cdot p$ we get 
\ball{d19}
1-\b n_0\cdot \b v_0 = \frac{\omega \e}{\omega_0\e_0}(1-\b n\cdot \b v)\,, 
\gal
accurate up to the terms of the order $m^2/\e^2$ and $M^2/\omega^2$. 

The transformation of the photon emission rate reads \cite{Tuchin:2013bda}
\ball{d21}
\frac{d\dot w}{d\Omega d\omega}= \frac{1}{\gamma^2(1+V\cos\theta)}\frac{d\dot w_0}{d\Omega_0 d\omega_0}=\frac{\omega \e_0}{\e \omega_0}\frac{d\dot w_0}{d\Omega_0 d\omega_0} \,,
\gal
where $\theta$ is angle between the photon momentum $\b k$ and the magnetic field, i.e.\ $\cos\theta = n_y$. In the last step we used \eq{d15} and \eq{d17}. $d\dot w_0$ in the right-hand-side of \eq{d21} is given by \eq{b77} and \eq{b78} with the replacements $\e\to \e_0$, $\omega\to \omega_0$ etc.

\section{Vector boson radiation by a plasma}\label{sec:f}

A system of electrically charged particles in thermal equilibrium in external magnetic field radiates vector bosons at the following rate per unit interval of vector boson energy $d\omega$ into a solid  angle $d\Omega$: 
\ball{f11}
\frac{dN}{dtd\Omega d\omega}= 2N_c\sum_{f}\int \frac{d\mathcal{V} d^3p}{(2\pi)^3} f(\e)[1-f(\e')]
\frac{d\dot w}{d\Omega d\omega}\,,
\gal
where the sum runs over all charged particle species in plasma, and $f(\e)$ are their distribution functions. Integration over the fermion momentum can be done using a Cartesian reference frame span by three unit vectors $ \b e_1, \b e_2, \b n$, such that vector $\b B$ lies in  plane span by $\b e_1, \b n$. In terms of the polar and azimuthal angles $\chi$ and $\psi$ we can write
\bal
&\b v= v(\cos\chi \, \b n+ \sin\chi \cos\psi \,\b e_1+\sin\chi \sin\psi \,\b e_2)\,,\label{f15}\\
& \b B= B(\cos\theta\, \b n_1+\sin\theta\, \b e_1)\,.\label{f16}
\gal
The element of the solid angle is $do= d\cos\chi\, d\psi$. In this reference frame
\bal
&p_y= \frac{\b p\cdot \b B}{B}= \e v(\cos\chi\cos\theta+\sin\chi \cos\psi \sin\theta)\,,\label{f19}\\
&k_y= \frac{\b k\cdot \b B}{B}=k \cos\theta\,,\label{f20}\\
& \b n\cdot \b v = v\cos\chi\,.\label{f21}
\gal

Fermions moving in plasma parallel to the magnetic field direction do not radiate due to the vanishing Lorentz force. Taking into account that at high energies fermions radiate mostly into a narrow cone with the opening angle $\chi \sim m/\e, M/\omega$ (see \eq{a21}), we conclude that vector boson radiation at angles $\theta\lesssim m/\e,M/\omega$ can be neglected. Thus, expanding at small $\chi$ we obtain from \eq{d17},\eq{f19}
\ball{f23}
\e_0\approx   \e \sin\theta\,,\quad \omega_0\approx \omega \sin\theta\,, \qquad \theta >\frac{m}{\e},\frac{M}{\omega}\,.
\gal
Omission of terms of order $m/\e$, $M\omega$ is  consistent with the accuracy of \eq{b77},\eq{b78}. The dependence of the integrand of \eq{f11} on the fermion direction specified by the angles $\chi$, $\psi$ comes only through \eq{d19}, viz.\ 
\ball{f25}  
1-\b n_0\cdot \b v_0= \frac{1}{\sin^2\theta}\left(1-\cos\chi +\frac{m^2}{2\e^2}\right)\,.
\gal
For this reason, integration over the quark momentum directions is similar to the one that led us from \eq{b73}, \eq{b74} to \eq{b81}, \eq{b82} (in the $K_0$ reference frame). 
Writing \eq{f11} as 
\bal
\frac{dN}{dt d\Omega d\omega}= &\frac{2N_c}{(2\pi)^3}\sum_{f}\int d\mathcal{V}\int_\omega^\infty d\e\, \e^2  f(\e)[1-f(\e')]
\sum_{\lambda=L,T} \int do\, \frac{d\dot w_\lambda}{d\Omega d\omega}
\label{f31}
\gal
and substituting \eq{d21}, \eq{b73}, \eq{b74} (with appropriate notation changes as described in \sec{sec:d}) and \eq{f23} we integrate first over $do$ and then over $\tau$ with the following result
\bal
\int do\, \frac{d\dot w_{T}}{d\Omega d\omega}= &-\frac{\alpha m^2}{\e^2}\sin^2\theta\left\{ \left( 1+\frac{M^2}{2\omega^2}\frac{\e^2+\e'^2}{m^2}\right)\int_{z_\theta}^\infty \text{Ai}(z')dz'\right. \nonumber\\
&\left. +(\sin\theta)^{2/3} \left( \frac{\e}{\e'}\right)^{1/3} \left( \frac{\omega_B}{\omega}\right)^{2/3}\frac{\e^2+\e'^2}{m^2} \text{Ai}'(z_\theta)
\right\}\label{f41}\,,\\
\int do \,\frac{d\dot w_{L}}{d\Omega d\omega}= &\frac{\alpha M^2 \e'}{\omega^2\e}\sin^2\theta\int_{z_\theta}^\infty \text{Ai}(z')dz' \,,\label{f42}
\gal
where 
\ball{f44}
z_\theta=(\sin\theta)^{-2/3}\left( \frac{\e}{\e'}\right)^{2/3} \left( \frac{\omega}{\omega_B}\right)^{2/3}\left( \frac{m^2}{\e^2\sin^2\theta}+\frac{M^2}{\omega^2}\frac{\e'}{\e}\right)\,.
\gal

If the magnetic field is a slow function of time and/or coordinates one can adopt an adiabatic approximation and integrate \eq{f41} and \eq{f42} over the time and space which yields 
the total vector boson multiplicity spectrum radiated into a unit solid angle. This is the formula that has been recently employed in \cite{Tuchin:2014pka} for the calculation of  the synchrotron radiation of real photons in heavy-ion collisions, which is one of the outstanding problems in the high energy nuclear physics   \cite{Shuryak:1978ij,Hwa:1985xg,Kapusta:1991qp,Baier:1991em,Arnold:2001ba,Arnold:2002ja,Turbide:2007mi,vanHees:2011vb}.

For practical applications in relativistic heavy-ion phenomenology it is customary to represent the bosom spectra as functions of  rapidity $y$  and transverse momentum $k_\bot$ with respect to the collision axis  $z$, in place of  energy $\omega$ and  emission angle $\theta$ with respect to the magnetic field.  Let $\alpha$ and $\phi$ be the polar and azimuthal angles of boson with respect to the collision axis. They are related to $\omega$ and $\theta$ as follows \cite{Tuchin:2012mf}:
\beql{coor2}
\omega= k_\bot \cosh y\,,\quad \cos\theta= \frac{\sin\phi}{\cosh y}\,.
\eeq
The differential boson multiplicity can be represented as
\beql{mult2}
\frac{dN}{dVdt\,d^2k_\bot dy}= \frac{dN}{dVdt\,\omega d\omega d\Omega}\,,
\eeq
where one should substitute \eq{coor2} in the right-hand-side of \eq{mult2}.

In deriving \eq{f31}--\eq{f44} we assumed that plasma is relativistic, i.e.\ that fermion energy satisfies $\e\sim T\gg m$. This condition must hold not only for the current mass $m$, but also for the temperature dependent contribution that fermions receive due to their interaction with the plasma. Evidently, this contribution must be small compared to the plasma temperature. This is true in a weakly coupled plasma, such as the electromagnetic plasma, because fermion mass receives a correction of order $gT\ll T$, where $g$ is the coupling constant. As far as the quark-gluon plasma is concerned, the coupling $g$ is not small at temperatures relevant in experiment. In practice, effective quark and gluon masses are treated as free parameters in models describing the quark-gluon plasma. Under such circumstances accuracy of the ultra-relativistic limit used to derive \eq{f31}--\eq{f44} depends on a particular model used to describe the plasma dynamics.

\section{Summary}\label{sec:con}

In this paper we used the quasi-classical method to derive the synchrotron radiation rate of massive vector bosons  including virtual photons. The main result is expressed in  formulas \eq{b77}--\eq{b84} that give the vector boson radiation rate by a relativistic electrically charged fermion. They describe spectrum and the angular distribution of ultra-relativistic vector bosons. Our analysis of the mass dependence of the synchrotron spectra revealed that with increase of $M$, spectra become increasingly monochromatic with energy $\omega_c$, given by \eq{b95}. A more detailed structure is shown in \fig{fig1} and \fig{fig2}.

Eqs.~\eq{b77}--\eq{b84}  can be directly applied to investigate the  space-time structure of magnetic field and its dynamics in experiments with intense laser beams. In view of possible applications 
in high energy nuclear physics and in astrophysics, we derived vector boson spectrum \eq{f31}--\eq{f44} radiated by a relativistic plasma.  These equations can be used, for example,  to evaluate a contribution of synchrotron radiation to the dilepton spectrum produced in relativistic heavy-ion collisions at $k_\bot>M$ and $y=0$ and compare with the experimental data reported in \cite{Adare:2009qk}.
 
 These and other applications deserve full consideration in separate publications.

\acknowledgments

This work  was supported in part by the U.S. Department of Energy under Grant No.\ DE-FG02-87ER40371.

\appendix
\section{Some useful integrals involving the Airy function $\text{Ai}(z)$}\label{appA}

In the following integrals $a$, $b$ are real numbers and $z=  a/(3b)^{1/3}$.
\begin{align}
&\frac{1}{2\pi}\int_{-\infty}^\infty  e^{-i[a\tau +b\tau^3]}d\tau= \frac{1}{(3b)^{1/3}}\,\text{Ai}(z)\,,\label{ap11}\\
&\frac{1}{2\pi i}\int_{-\infty}^\infty  \tau e^{-i[a\tau +b\tau^3]}d\tau= \frac{1}{(3b)^{2/3}}\,\text{Ai}'(z)\,,\label{ap12}\\
&\frac{1}{2\pi}\int_{-\infty}^\infty  \tau^2 e^{-i[a\tau +b\tau^3]}d\tau=-\frac{ z}{3b}\,\text{Ai}(z) \,,\label{ap13}\\
&\frac{1}{2\pi i}\int_{-\infty}^\infty  \frac{1}{\tau} e^{-i[a\tau +b\tau^3]}d\tau=\int_z^\infty \text{Ai}(z') \,dz'\,,\label{ap14}
\end{align}


\end{document}